\documentclass[preprint,sort&compress,times,12pt]{elsarticle}
\usepackage{amsmath,graphicx}
\newcommand{\braket}[1]{\langle #1 \rangle}
\begin{document}
\title{Understanding agent-based models of financial markets: a
bottom-up approach based on order parameters and phase diagrams}
\author[pap]{Ribin Lye}
\author[pap]{James Peng Lung Tan}
\author[pap]{Siew Ann Cheong\corref{cor1}}
\ead{cheongsa@ntu.edu.sg}
\cortext[cor1]{Corresponding author}
\address[pap]{Division of Physics and Applied Physics, 
School of Physical and Mathematical Sciences,
Nanyang Technological University,
21 Nanyang Link, Singapore 637371,
Republic of Singapore}

\begin{abstract}

We describe a bottom-up framework, based on the identification of appropriate
order parameters and determination of phase diagrams, for understanding
progressively refined agent-based models and simulations of financial markets.
We illustrate this framework by starting with a deterministic toy model, whereby
$N$ independent traders buy and sell $M$ stocks through an order book that acts
as a clearing house.  The price of a stock increases whenever it is bought and
decreases whenever it is sold.  Price changes are updated by the order book
before the next transaction takes place.  In this deterministic model, all
traders based their buy decisions on a call utility function, and all their sell
decisions on a put utility function.  We then make the agent-based model more
realistic, by either having a fraction $f_b$ of traders buy a random stock on
offer, or a fraction $f_s$ of traders sell a random stock in their portfolio.
Based on our simulations, we find that it is possible to identify useful order
parameters from the steady-state price distributions of all three models.  Using
these order parameters as a guide, we find three phases: (i) the dead market;
(ii) the boom market; and (iii) the jammed market in the the phase diagram of
the deterministic model.  Comparing the phase diagrams of the stochastic models
against that of the deterministic model, we realize that the primary effect of
stochasticity is to eliminate the dead market phase.

\end{abstract}


\maketitle

\section{Motivation}

Economies and financial markets are complex systems described by a large number
of variables, which are in turn influenced by an even larger number of factors
or players.  To understand how these complex systems evolve in time, an approach
that has been very fruitful thus far is to consider the dynamics of a small
number of aggregate collections of homogeneous variables.  In this approach, the
dependences of aggregate averages on other aggregate averages, and on time, are
modeled by coupled systems of ordinary or partial differential equations.  While
this equation-based approach has been able to provide rigorous theorems, and
generate much economic insights, it is fundamentally mean field in character, in
that variances and higher-order statistical moments of the aggregates are
assumed to be slaves of the averages, and have no independent dynamics of their
own.  In many important and interesting situations in the real world, this
assumption is indeed valid, because the number of variables in each homogeneous
aggregate is large, and the central limit theorem applies.

However, in many other interesting real-world situations, statistical
fluctuations can become an important driver in the time evolution of economies
and financial markets.  When this happens, the variances and higher statistical
moments of the aggregates become large, and their dynamics cannot be deduced
from those of the averages alone.  This is where agent-based models and
simulations become invaluable as a tool for understanding the dynamics of the
economic or financial system as a whole.  Since the pioneering work of Arthur
and co-workers \cite{Palmer1994PhysicaD75p264}, there has been rapidly growing
interest in the use of agent-based simulations as a computational platform for
performing `controlled experiments' in an economic setting
\cite{Epstein1999Complexity4p41, Marsili2007EurPhysJB55p169,
Farmer2002IndCorpChange11p895}.  This has culminated in several reviews
\cite{LeBaron2000JEconDynCont24p679, LeBaron2001QuantFin1p254,
Tesfatsion2002ArtificialLife8p55} and monographs \cite{HCE2006,
Levy2000MicroSimulFinMkt, Mantegna2000IntroEconophysics} on the subject.  In
general, economists have taken a top-down approach to agent-based modeling,
implementing neo-classical economic axioms, which are then systematically
relaxed to incorporate the effects of heterogenuity \cite{Lux1999Nature297p498,
Lux2000IntJTheorApplFin3p675}, long-term memory
\cite{Raberto2001PhysicaA299p319}, and learning
\cite{LeBaron1999JEconDynCont23p1487}.  While this line of agent-based research
has been able to give stylized and qualitative results, it is generally very
difficult to draw quantitative, and deeper inferences and insights, because of
large statistical fluctuations within the simulations.

In contrast, highly simplified models have been the choice of physicists
\cite{Bak1997PhysicaA246p430, Caldarelli1997EuroPhysLett40p479,
Sato1998PhysicaA250p231, Chowdhury1999EurPhysJB8p477, Cont2000MacroEconDyn4p170,
Maslov2000PhysicaA278p571, Iori2002JEconBehavOrg49p269}, because such models are
easy to understand in quantitative terms.  In this paper, we propose a bottom-up
framework for going between the two agent-based modeling approaches.  We start
by describing in Section \ref{sect:models} a highly simplistic agent-based
model, whose purpose is to illustrate the ideas behind the model differentiation
framework, rather than to accurately describe the real world.  After working out
its phase diagram in Section \ref{sect:results}, we refine the model by
introducing a stochastic component in the trading strategies of the agents.  By
examining how the phase diagram is modified by the additional rules, we report
insights gained from our preliminary study.  In Section \ref{sect:summary}, we
discuss how we can refine the model progressively to make its dynamics more and
more realistic, and ultimately develop a picture on the hierarchy of
complexities that emerge at various levels of realism in the models.  Our goal
is to eventually be able to identify robust market behaviours, which depend on
the gross structure of the models but not the details, and also fragile market
behaviours, which depends sensitively on certain model details, and hence are
expected to appear only rarely in the real world.

\section{Deterministic Model}
\label{sect:models}

To illustrate the basic framework in our bottom-up approach to understanding
agent-based models, we start from a highly simplified model of a financial
market, which consists of $N$ traders, $M$ stocks, and an order book.  At the
start of each simulation, we assign each of the $M$ stocks a random price $0 < p
\leq p_{\max}$, and a zero initial price change $\Delta p = 0$.  A random
initial offer $0 < q \leq q_{\max}$ for each stock is also made available on the
order book, which plays the role of a clearing house.  We then assign to each of
the $N$ traders a random initial capital $c = c_{\max}$, as well as a random
portfolio that excludes short positions.  In this simple model, we assume that
the $N$ traders do not directly interact with each other, but carry out
transactions only through the order book.  At each time step, the $N$ traders
will trade in a randomized sequence.  Each trader will buy one stock offered by
the order book, and then sell one stock in his or her portfolio. 

When it is trader $i$'s turn to trade, he or she will evaluate the utilities
$(u_1, \dots, u_M)$ of the $M$ stocks on offer in the order book, based on the
\emph{call function}
\begin{equation}
u_s = \frac{\alpha}{p_s} + \Delta p_s,
\end{equation}
where $p_s$ is the price of stock $s$, and $\Delta p_s$ is the last price change
of stock $s$.  Trader $i$ then buys however many units he or she can afford of
the stock $s^*$ having the maximum utility.  After this transaction, the order
book increases the price $p_{s^*}$ of stock $s^*$ by one unit, and sets $\Delta
p_{s^*} = +1$.  Once this price adjustment is completed, trader $i$ evaluate the
utilities $(v_1, \dots, v_M)$ of all $M$ stocks, using the \emph{put function}
\begin{equation}
v_s = p_s - \beta \Delta p_s.
\end{equation}
If stock $s^{**}$ is found to have the maximum utility, trader $i$
will sell all of stock $s^{**}$ in his or her portfolio.  The order
book then decreases the price $p_{s^{**}}$ of stock $s^{**}$ by one
unit, and sets $\Delta p_{s^{**}} = -1$ before the next trader trades.

In this model, we introduce two utility functions $u(p, \Delta p)$ and $v(p,
\Delta p)$ to accommodate potential asymmetry between seller and buyer trading
strategies.  In these two utility functions, we also attempt to incorporate both
fundamentalist and chartist tendencies.  Given two stocks in the order book with
the same positive long-term ratings, we expect fundamentalist traders to always
buy the cheaper of the two.  This tendency is accounted for in the call function
by the price term $\alpha/p$, which decreases with increasing price $p$.
Chartist traders, on the other hand, will only buy an appreciating stock.  Hence
our choice of the price change term $\Delta p$ in the call function.   We also
assume that stocks are merely tools to increase capital assets, and traders
attach no further value to them.  Therefore, for the purpose of generating cash
flow, we expect a fundamentalist trader will always sell his or her most
expensive stock.  This tendency is modeled by the price term $p$ in the put
function.  A chartist trader, whose trading decisions are based entirely on
price changes, will only sell depreciating stocks.  Hence our choice of the
price change term $-\beta \Delta p$ in the put function.  The traders in our
simulations, who have no memory (and thus no capacity to learn), exhibit these
fundamentalist and chartist tendencies to different extents, depending on the
two independent parameters $\alpha$ and $\beta$.  We vary $\alpha$ and $\beta$
to determine the phase diagram of this model, whereby all traders behave
rationally, and based their trading decisions on two deterministic utility
functions.

\section{Order Parameters and Phase Diagram}
\label{sect:results}

All our simulations were done with $N = 10,000$, $M = 1000$, $c_{\max} = 100$,
$p_{\max} = 100$, and $q_{\max} = 100$.  We chose these fixed parameters so that
our system of agents resemble, at a very gross level, small markets like the
Singapore Exchange (SGX), on which fewer than a thousand stocks are listed,
attracting about 10,000 active traders.  Assuming price movements are quantized
on the level of S\$0.05, the maximum stock price $p_{\max}$ that we used to
generate the initial distribution of stock prices, as well as the capital
$c_{\max}$ each trader is initially endowed with, both corresponds to an actual
level of S\$5.00.  Our simulated market is thus a penny-stock market initially,
played only by retail traders.  

\begin{figure}[htbp]
\centering
\includegraphics[scale=0.65]{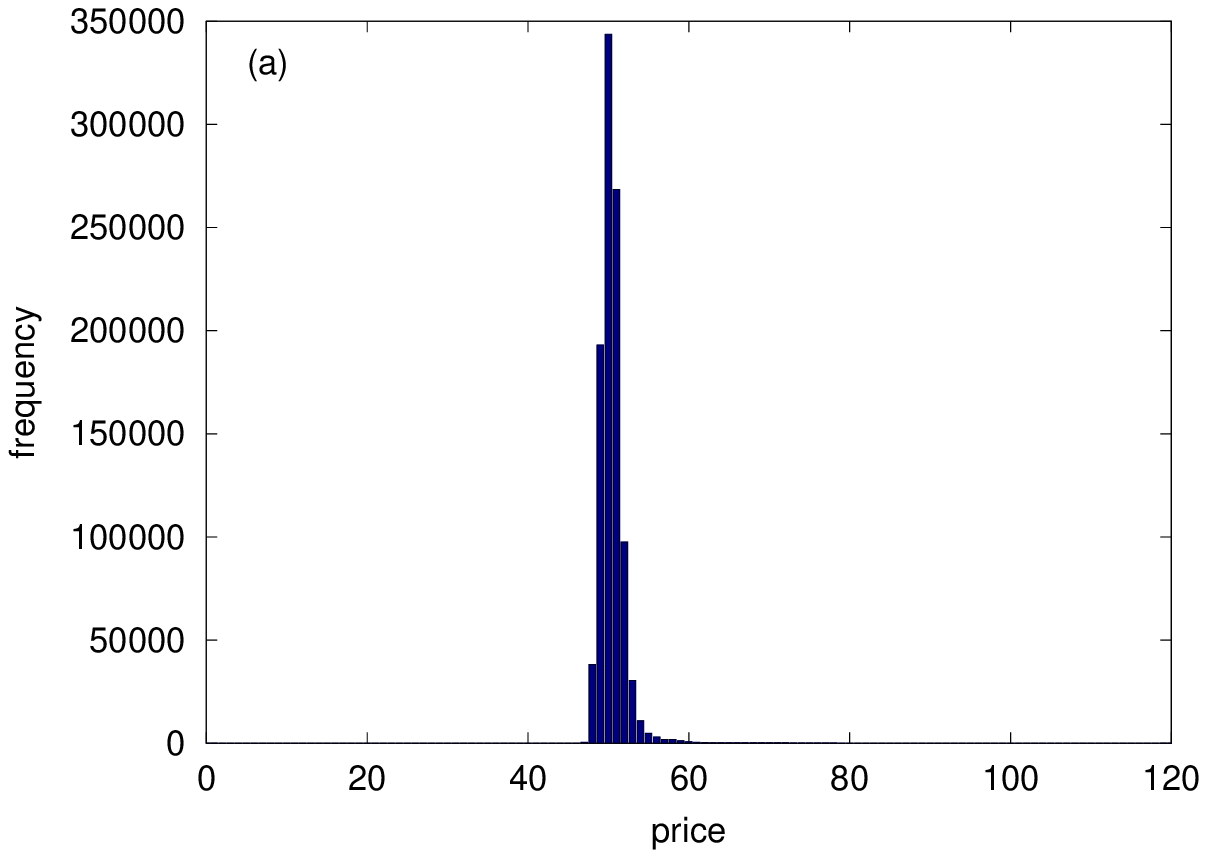}

\includegraphics[scale=0.65]{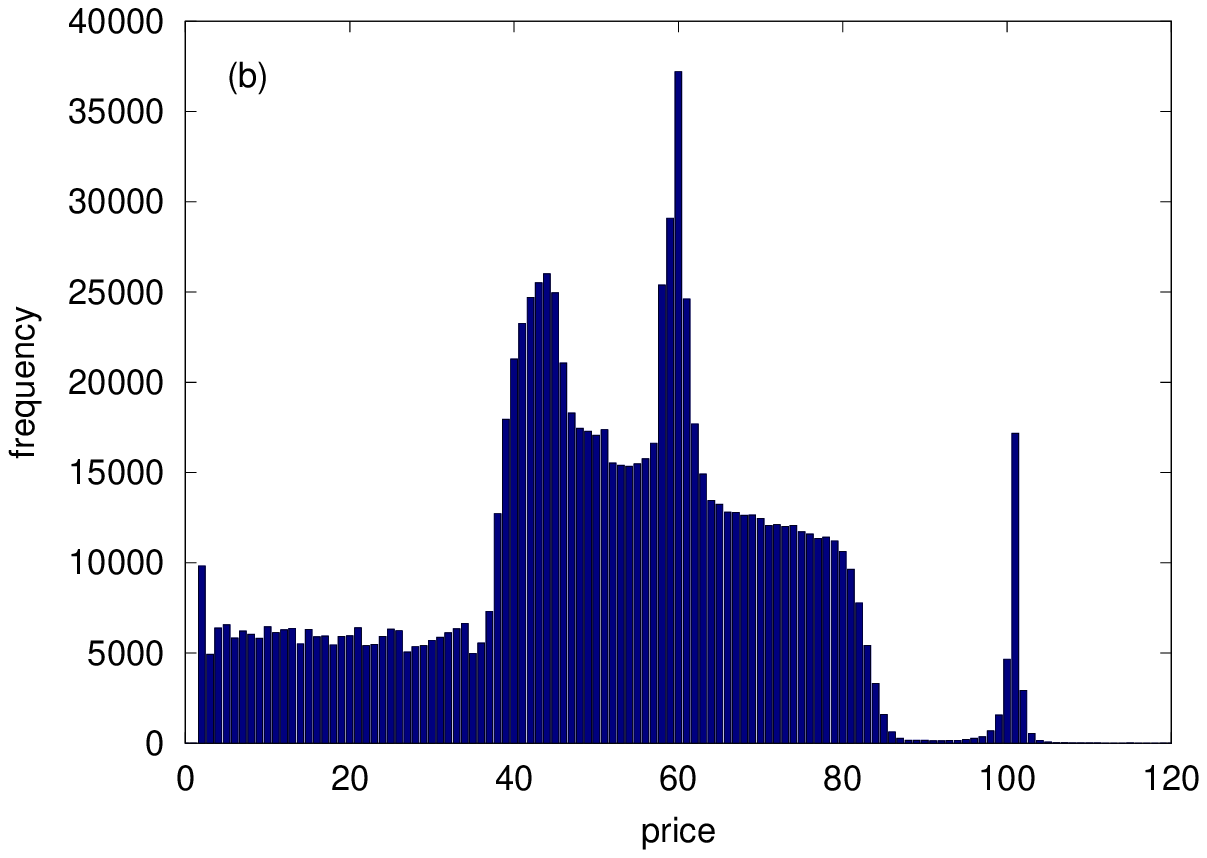}

\includegraphics[scale=0.65]{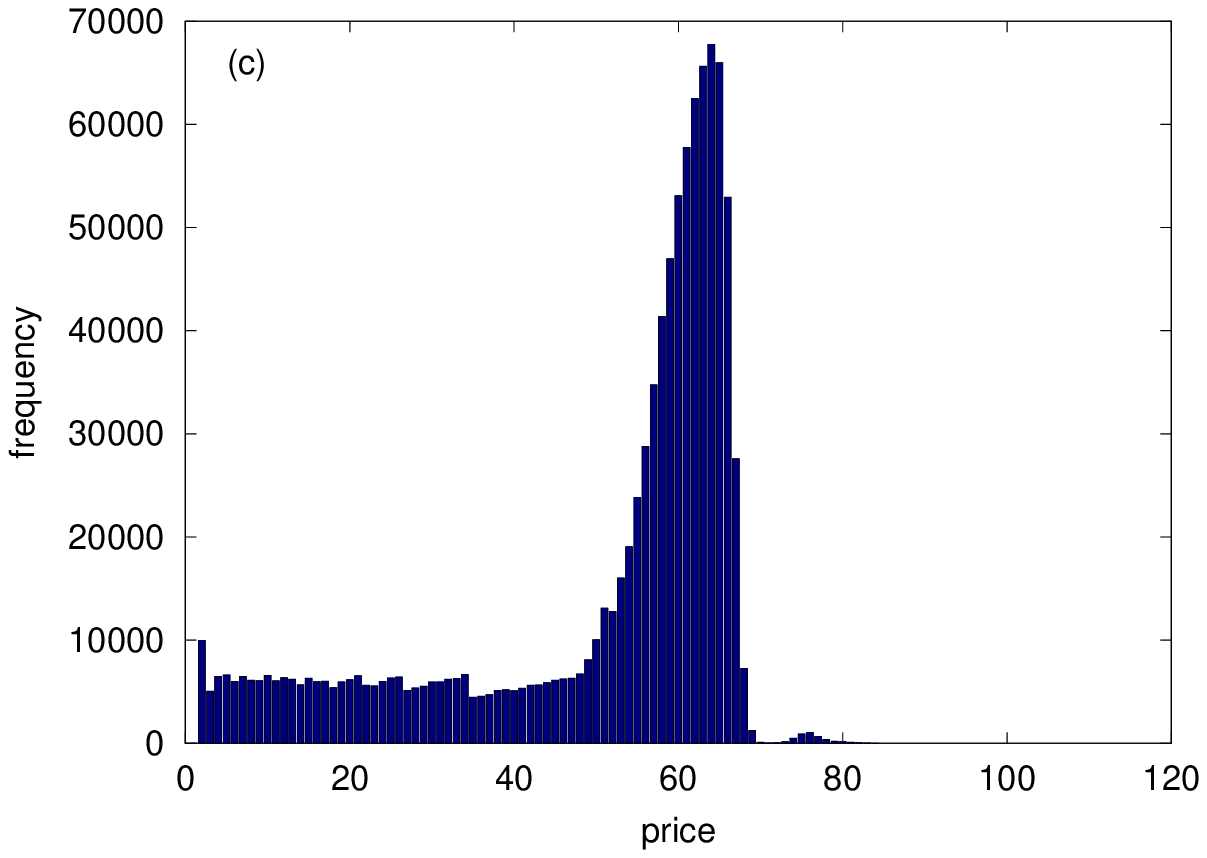}
\caption{Steady-state price distributions for (a) $\alpha = 10$, $\beta = 10$,
(b) $\alpha = 100$, $\beta = 100$, (c) $\alpha = 100$, $\beta = 10$.  Each
distribution is accumulated over 1000 simulations.} 
\label{fig:pricedist}
\end{figure}

We ran each simulation up to 100 time steps, and examine the distribution of
prices within all $M = 1000$ stocks.  We find that for most parameter values
$(\alpha, \beta)$, the initial uniform distribution of prices evolves into a
steady-state distribution within 5--10 time steps.  After a thorough examination
of a large region of the $\alpha$-$\beta$ parameter space, we identified two
robust features in the steady-state price distribution: (i) a uniform
sub-distribution of below-average stock prices; and (ii) a gaussian
sub-distribution of above-average stock prices.  Based on features of these two
sub-distributions, we identify three phases, I, II and III, for the
deterministic model.  In Figure \ref{fig:pricedist}, we show typical price
distributions in phases I, II, and III.  In phase I (Figure
\ref{fig:pricedist}(a)), the steady-state price distribution is very narrowly
distributed about $\bar{p} = 50$.  Both the uniform sub-distribution of
below-average prices and the gaussian sub-distribution of above-average prices
are absent.  The uniform sub-distribution of below-average prices can be found
in phases II (Figure \ref{fig:pricedist}(b)) and III (Figure
\ref{fig:pricedist}(c)), which are distinguished by the gaussian
sub-distribution of above-average prices.  In phase II, this gaussian
sub-distribution is strong and narrow, with a peak position that increases with
$\alpha$, whereas in phase III, this gaussian sub-distribution is weak and
broad, with a peak position that is independent of $\alpha$.

\begin{figure}[htbp]
\centering
\includegraphics[scale=0.35]{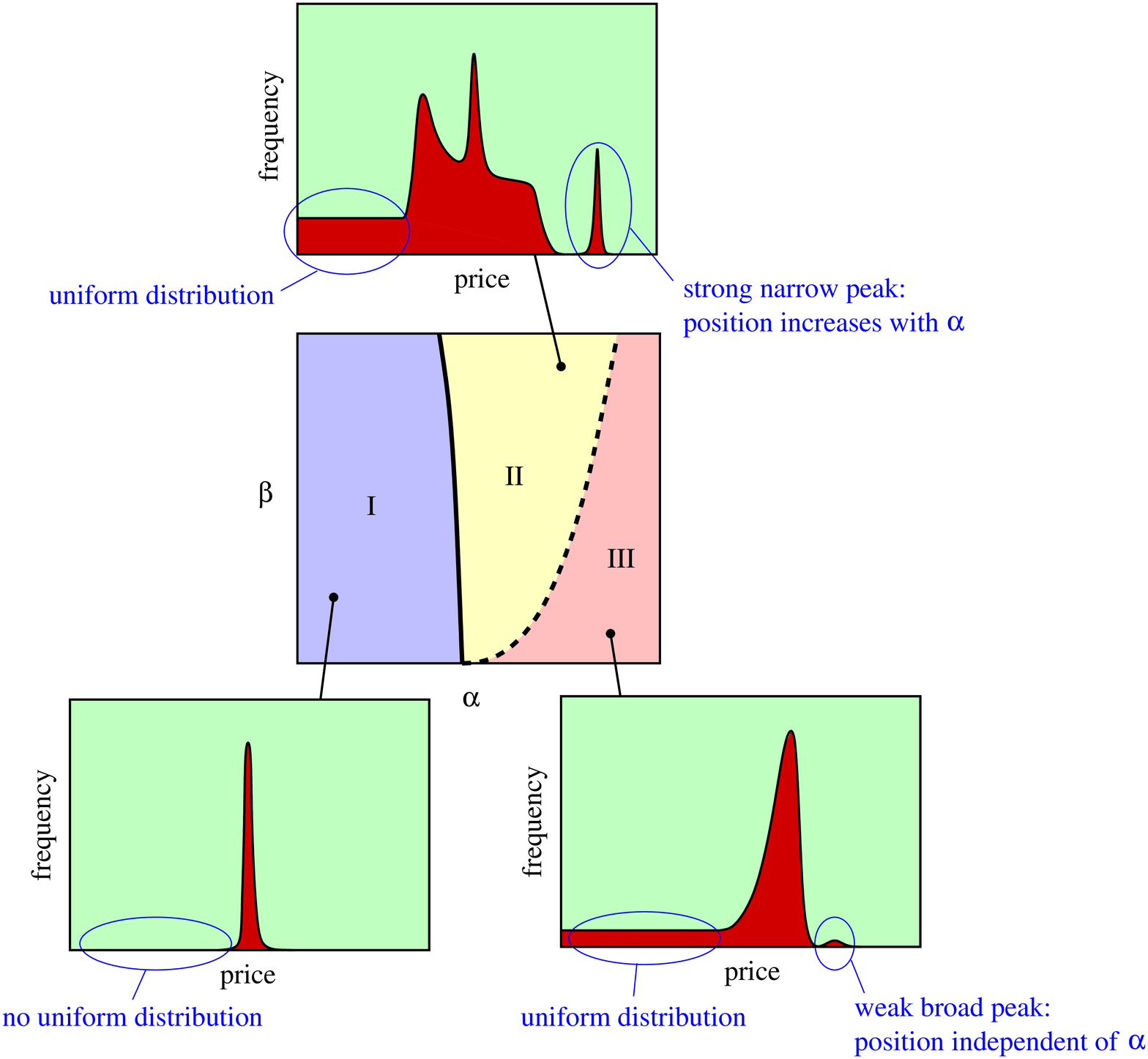}
\caption{Schematic phase diagram of the deterministic model, showing the three
phases (I, II, and III), as well as caricatures of their steady-state price
distributions.  In this diagram, a solid line denotes a line of second-order
phase transitions, whereas a dashed line denotes a line of crossovers.}
\label{fig:Dphase}
\end{figure}

Our exploration of the $\alpha$-$\beta$ parameter space led us to sketch the
phase diagram shown in Figure \ref{fig:Dphase}.  Since our toy model is not
intended to be a realistic model of any financial market, there is no real need
to understand the phase diagram in detail.  Nevertheless, the steady-state price
distribution in Phase I is easy to understand: the buying of cheap stocks drive
their prices up, and the selling of expensive stocks drive their prices down.
This trading dynamics, which is essentially anti-diffusion in nature, eventually
produces the sharp gaussian peak seen in the Phase I steady-state price
distribution.  We call Phase I a \emph{dead market}, because there is very
little trading activity in the steady state.  While we do not understand all the
features in the steady-state price distributions of Phases II and III, we do
realize that in Phase II, traders can get increasing returns by `latching' on to
the sub-distribution of above-average stock prices.  We therefore call this
phase the \emph{boom market}.  In Phase III, this sub-distribution of
above-average stock prices does not move as we increase $\alpha$, and so we call
this phase the \emph{jammed market}. 

Now, the idea of studying phase transitions in economic systems is not
new.  LeBaron and co-workers recognize the existence of distinct
economic phases from their early work on agent-based modeling
\cite{LeBaron1999JEconDynCont23p1487}.  More recently, Giardina and
co-workers sketched the phase diagram of a sophisticated agent-based
model to explain bubbles, crashes, and intermittent time series
dynamics observed in real markets \cite{Giardina2003EurPhysJB31p421,
Giardina2003PhysicaA324p6}, while Moukarzel and co-workers constructed
the phase diagram of a economy-scale agent-based model to explain the
phenomenon of wealth condensation
\cite{Moukarzel2007EurPhysJSpecTop143p75}.  Since our goal is to
eventually be able to compare phase diagrams across progressively
refined models, it is important for us to be able to draw stronger and
more quantitative inferences.  To accomplish this, its is necessary to
introduce \emph{order parameters} to characterize the observed phases,
like what Gauvin and co-workers did for their agent-based model to
study sociological segregation \cite{Gauvin2009socph09034694}.  

\begin{figure}[htbp]
\centering
\includegraphics[scale=0.45,clip=true]{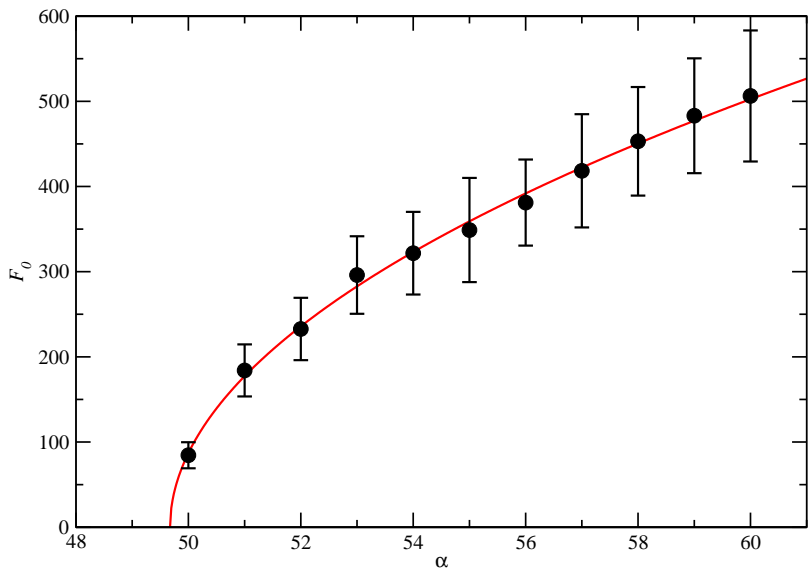}
\caption{Height $F_0$ (solid circles) of the uniform sub-distribution across the
I-II phase transition line, plotted as a function of the parameter $\alpha$, for
$\beta = 10$.  The error bar for each data point is estimated from the uniform
sub-distributions of 1000 simulations.  The red solid curve is the power-law
function $F_0 = G_0\, (\alpha - \alpha_c)^{\gamma}$, with $G_0 = 153.565$,
$\alpha_c = 49.6731$, and $\gamma = 0.507745$, that best fits the data points.}
\label{fig:Dphaseexp} 
\end{figure}

In general, a useful order parameter must be non-zero in one phase, and zero in
the rest of the phases.  From Figure \ref{fig:Dphase}, it is clear that the
height $F_0$ of the uniform sub-distribution is a good order parameter for
distinguishing between Phase I and Phases II/III.  In Figure
\ref{fig:Dphaseexp}, we show the dependence of $F_0$ on $\alpha$, for $\beta =
10$ kept fixed.  From the theory of critical phenomena in statistical physics,
we know that statistical fluctuations in a critical system are equally strong at
all scales, and thus order parameters are typically scale-free power laws in the
vicinity of critical points.  Indeed, we find a power law of the form $F_0 = G_0
(\alpha - \alpha_c)^{\gamma}$ fits the graph in Figure \ref{fig:Dphaseexp} very
well, with best-fit values $G_0 = 153.565$, $\alpha_c = 49.6731$, and $\gamma =
0.507745$.  With such a scaling form, the order parameter $F_0$ is singular at
$\alpha = \alpha_c$, and the line separating Phase I and II is thus a line of
true phase transitions (denoted by a solid curve in Figure \ref{fig:Dphase}).
In contrast, if we use the peak position $\alpha_0$ of gaussian sub-distribution
of above-average stock prices as an `order parameter' to distinguish Phase II
from Phase III, we find $\alpha_0$ changing continuously with $\alpha$ as we
move from Phase II into Phase III.  Therefore, as far as we can tell, there is
no true phase transitions between Phase II and Phase III, and we use a dashed
curve to indicate the \emph{crossover} from Phase II to Phase III.

Another important result that emerges from the study of critical phenomena is
the existence of universality classes of models, each characterized by a set of
universal \emph{critical exponents}.  For different values of $\beta$, we find
that the fitted values of $\gamma$ are nearly the same.  This suggests that
$\gamma$ is a universal critical exponent.  Moreover, these fitted values of
$\gamma$ are all very close to the universal exponent value of $\frac{1}{2}$,
which is associated with ferromagnetic phase transitions in the mean-field limit
(a limit attained by the Ising spin lattice model in infinite dimensions)
\cite{Binney1992}.  This observation is surprising, since traders in our
agent-based model do not interact directly with each other, unlike explicit
Ising-like interactions between traders in the highly-simplified market models
considered by Johansen and co-workers \cite{Johansen1999EurPhysJB9p167,
Johansen2000IntJTheorApplFin3p219}.

We believe that through their interactions with the order book, traders
experience a \emph{retarded} Ising-like interaction with other traders.
Retarded interactions offer surprises in many other systems.  For example, in
conventional superconductors, electrostatically repelling electrons experience a
retarded, effectively attractive interaction mediated by the ionic lattice, and
thereon condensed into Cooper pairs.  To test this hypothesis, we define a spin
variable $\sigma_i$ for trader $i$, such that
\begin{equation}
\sigma_i = \begin{cases}
+1, & \text{if $i$ buys}; \\
-1, & \text{if $i$ sells}.
\end{cases}
\end{equation}
In our toy market model, every agent will buy once and sell once every time
step.  Therefore, we cannot detect any ferromagnetic or antiferromagnetic
transition by plotting $\braket{\sigma_i}$ as a function of the parameter
$\alpha$.

Instead, let us plot the spin-spin correlation $\braket{\sigma_i\sigma_j}$ as a
function of $\alpha$.  To define this spin-spin correlation properly, let us
start with the definition
\begin{equation}
\sigma_{ist} = \begin{cases}
+1, & \text{if $i$ buys stock $s$ at time $t$}; \\
-1, & \text{if $i$ sells stock $s$ at time $t$}. \end{cases}
\end{equation}
We then compute the product
\begin{equation}
\sigma_{ist}\sigma_{jst} = \begin{cases}
+1, & \text{if $i$ and $j$ both buy or both sell stock $s$ at time $t$}; \\
-1, & \text{if $i$ buys stock $s$, while $j$ buys stock $s$ at time $t$, or vice
versa}, \end{cases}
\end{equation}
and its time average
\begin{equation}
\overline{\sigma_{ist}\sigma_{jst}} = \frac{1}{T}\sum_{t=1}^T
\sigma_{ist}\sigma_{jst}.
\end{equation}
If this time average is zero, then $\sigma_{is}$ and $\sigma_{js}$ are not
correlated.

However, since traders $i$ and $j$ can only buy and sell one stock each time
step, $\overline{\sigma_{ist}\sigma_{jst}}$ cannot be large.  To get a better
signal-to-noise ratio, we define the ensemble average
\begin{equation}
\braket{\sigma_i \sigma_j} = \frac{1}{M}
\sum_{s=1}^M \overline{\sigma_{ist}\sigma_{jst}}.
\end{equation}
This quantity is large only if traders $i$ and $j$ are always buying and selling
the same stocks.  Finally, we define the susceptibility
\begin{equation}
\chi = \sum_{i < j} \braket{\sigma_i \sigma_j},
\end{equation}
which will be large only if there are strong trading-induced correlations within
the toy financial market.

\begin{figure}[htb]
\centering
\includegraphics[scale=0.5,clip=true]{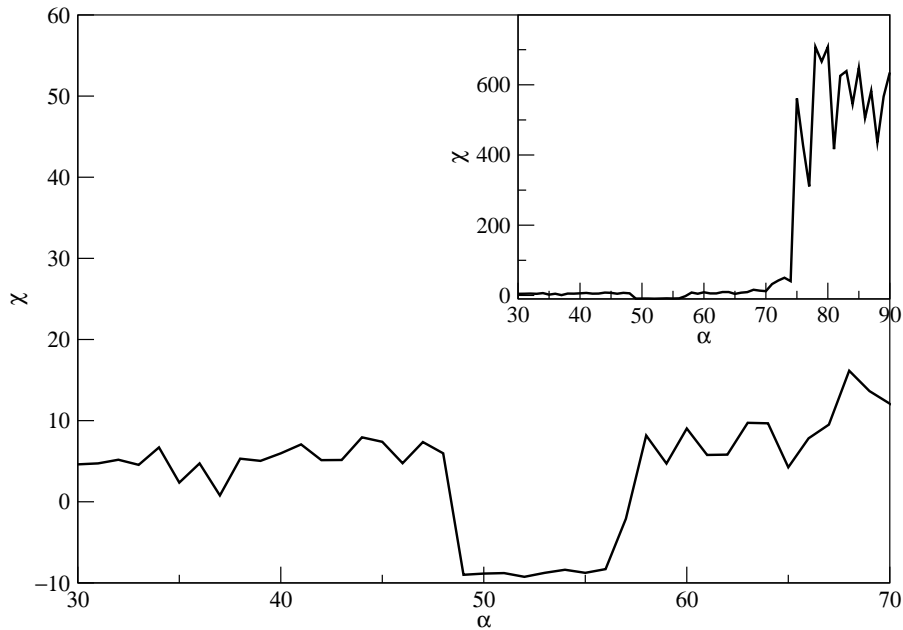}
\caption{The susceptibility $\chi = \sum_{i < j} \braket{\sigma_i \sigma_j}$ of
the toy market model, for $30 \leq \alpha \leq 70$ and $\beta = 10$.  For this
value of $\beta$, there is a second-order phase transition from Phase I to Phase
II at $\alpha_c \approx 50$. (Inset) The susceptibility $\chi$ over the broader
range of $30 \leq \alpha \leq 90$.  Note the sudden jump in $\chi$ close to the
crossover from Phase II to Phase III at $\alpha_0 \approx 78$ for $\beta = 10$.}
\label{fig:chi}
\end{figure}

In each time step $t$, the traders buy one stock each in random order, and after
all traders have bought a stock each, they then sell one stock each in a
different random order.  Therefore, in computing $\chi$, we add one to $\chi$
for every pair of traders buying the same stock during the buy half-cycle.  We
then subtract one from $\chi$ for every pair $(i, j)$, if $j$ had in the sell
half-cycle sold the stock that $i$ bought in the buy half-cycle, before adding
one to $\chi$ for every pair of traders selling the same stock during the sell
half-cycle.  Before this is repeated for the next time step $t+1$, we compare
the buy half-cycle for $t+1$ and the sell half-cycle for $t$, and subtract one
from every pair $(i, j)$, if $j$ had bought the stock that $i$ sold.  Finally,
we divide this accumulated result by $T$ and $M$.  For one particular sequence
of simulations, with $30 \leq \alpha \leq 90$, we find the results shown in
Figure \ref{fig:chi}.

As we can see from Figure \ref{fig:chi}, $\chi$ is non-zero even for $\alpha$
far away from $\alpha_c$.  We suspect this is due to the fact that $N = 10,000$
agents are trading $M = 1000$ shares, and therefore there will always be a
positive statistical background of traders buying or selling the same stocks.
There is also a negative statistical background of pairs of traders, one buying
a stock, and the other selling the same stock.  However, the positive background
comes about through the choice of a pair from $N$ items, but the negative
background comes about through the choice of a pair from $2N$ items, thus the
positive background prevails.  This means that the strong dip in $\chi$ around
$\alpha_c$ must be due to strong antiferromagnetic correlations mediated by the
order book.  Interestingly, we also see in $\chi$ a very pronounced signature of
the crossover from Phase II to Phase III, which occurs at $\alpha_0 \approx 78$
for $\beta = 10$.

\section{Stochastic Sell and Buy Models}
\label{sect:stochastic}

In real markets, traders do not always behave rationally.  To introduce some
stochasticity into the trading, we can have a fraction $f$ of the traders ignore
the utility functions, and always trade randomly.  Alternatively, we can have
\emph{all} traders trading randomly a fraction $f$ of the time.  We implement
the latter in our simulations, because there is no need to track the stochastic
traders individually.  If we keep the buy decision deterministic, and make each
trader sell a random stock in his or her portfolio a fraction $f_s$ of the time,
we end up with a \emph{stochastic sell model}.  Conversely, if we keep the sell
decision deterministic, and make each trader buy a random stock from the order
book a fraction $f_b$ of the time, we end up with a \emph{stochastic buy model}.
Both are slightly more realistic refinements of the deterministic model.  It is
also possible to introduce stochasticity into both the buy and sell decisions.
This is a more complex model that we intend to study only after we understand
the phase diagrams of the deterministic model, and the two simpler stochastic
models.

\begin{figure}[htbp]
\centering
\includegraphics[scale=0.5,clip=true]{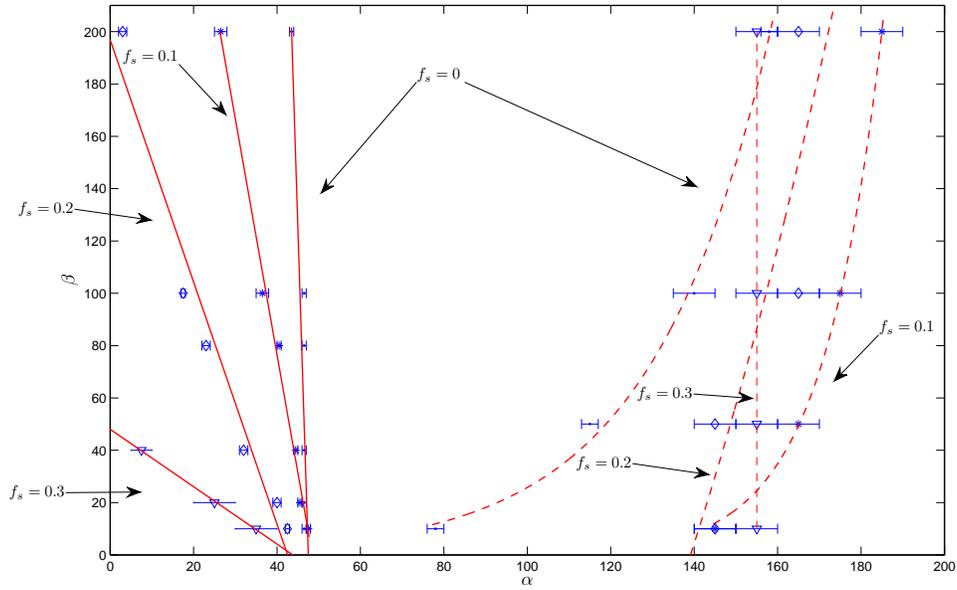}
\caption{Phase diagram of the stochastic sell model, showing how the phase
boundaries depend on the stochastic fraction $f_s$.  Qualitatively, we see that
as $f_s$ is increased, Phase II (boom market) expands into Phase I (dead
market).  However, as the stochastic fraction was increased from $f_s = 0$
through $f_s = 0.1$, Phase II (boom market) first expanded into Phase III
(jammed market), before retreating from $f_s = 0.1$ to $f_s = 0.2$, and more or
less staying put from $f_s = 0.2$ to $f_s = 0.3$.  In this plot, the true phase
transition line separating Phase I and Phase II are fitted to straight lines of
the form $\beta = m\alpha + c$, and shown as solid lines, while the line of
crossover from Phase II to Phase III are fitted to a mixture of exponentials and
polynomials, and shown as dashed lines.  The error bar $\Delta\alpha_c$ for each
data point $(\bar{\alpha}_c, \beta)$ is estimated by first breaking the 1000
simulations for each $(\alpha, \beta)$ into 10 sets of 100 simulations.  We then
performed power-law fits on these 10 sets of $F_0(\alpha)$, to obtain 10
$\alpha_c$.  Finally, we calculate $\bar{\alpha}_c$ and $\Delta\alpha_c$.}
\label{fig:Sphase}
\end{figure}

\subsection{Stochastic Sell Model}

\begin{figure}[htb]
\centering
\includegraphics[scale=0.5]{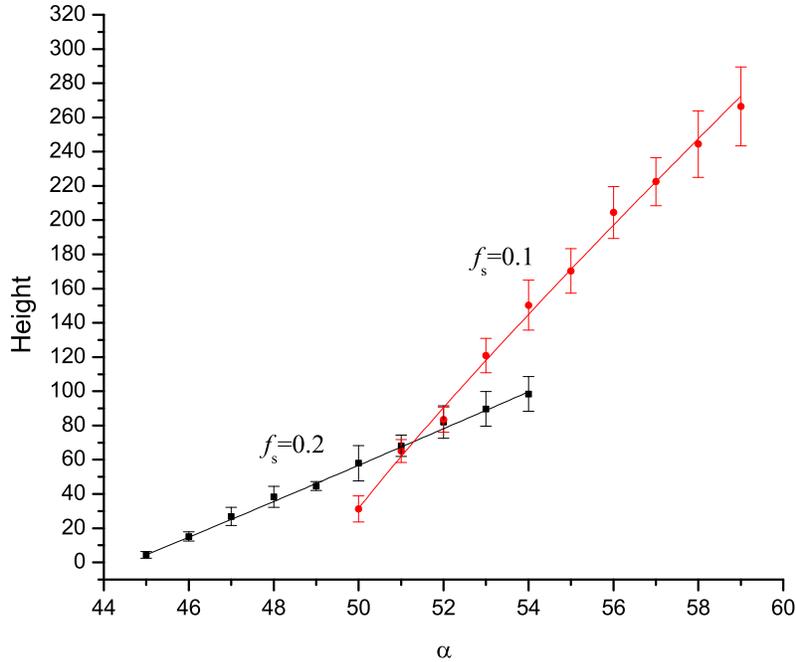}
\caption{Height $F_0$ (solid circles) of the uniform sub-distribution across the
I-II phase transition line, plotted as a function of the parameter $\alpha$, for
$\beta = 10$, and $f_s = 0.1$ (red) as well as $f_s = 0.2$ (black).  The error
bar for each data point is estimated from the uniform sub-distributions of 1000
simulations.  The solid curves are the best power-law fits of the $F_0 = G_0
(\alpha - \alpha_c)^{\gamma}$, where we find $G_0 = 34.9 \pm 5.9$, $\alpha_c =
49.10 \pm 0.26$, $\gamma = 0.896 \pm 0.070$ for $f_s = 0.1$, and $G_0 = 9.9 \pm
1.0$, $\alpha_c = 44.53 \pm 0.11$, $\gamma = 1.026 \pm 0.049$ for $f_s = 0.2$.}
\label{fig:Sphaseexp}
\end{figure}

Having developed a good understanding of the deterministic model ($f_s = 0$), we
move on to study the phase diagram of the stochastic sell model, for $f_s =
0.1$, $0.2$, and $0.3$.  In Phases I, II, and III, the typical steady-state
price distributions are very similar to those seen in the deterministic model.
Also, the order parameters identified for the deterministic model remained good
order parameters for the stochastic sell model.  After scanning over a broad
region in the parameter space, we did not find any new emergent phases, and thus
focussed on how the boundaries of the existing phases change as $f_s$ is varied,
as shown in Figure \ref{fig:Sphase}.  We find that, as $f_s$ is increased, Phase
II (boom market) expands into the region occupied by Phase I (dead market).  A
simple extrapolation from the graph of the slope of the phase transition line as
a function of $f_s$ suggests that the dead market phase will completely
disappear by $f_s = 0.4$.  This is reassuring, because the dead market phase is
not realistic, and we can be confident it will not appear in real markets that
are sufficiently noisy.  We also checked the character of the I-II phase
transition for $f_s = 0.1$ and $f_s = 0.2$, as shown in Figure
\ref{fig:Sphaseexp}, and find it becoming less abrupt with increasing $f_s$,
until it also becomes a line of crossovers for $f_s \gtrsim 0.2$.

We also find in Figure \ref{fig:Sphase} an interesting behaviour of the line of
crossovers: Phase II (boom market) expands at first into Phase III (jammed
market) when we go from $f_s = 0$ to $f_s = 0.1$, but retreated as we go from
$f_s = 0.1$ to $f_s = 0.2$.  This suggests that there is an optimum level of
trading stochasticity on the sell side, that can drive a boom of the gaussian
sub-distribution of above-average stock prices, before this sub-distribution
gets stuck in Phase III.  
%We intend to look into this phenomenon more carefully in the near future.

\subsection{Stochastic Buy Model}

\begin{figure}[htb]
\centering
\includegraphics[scale=0.45]{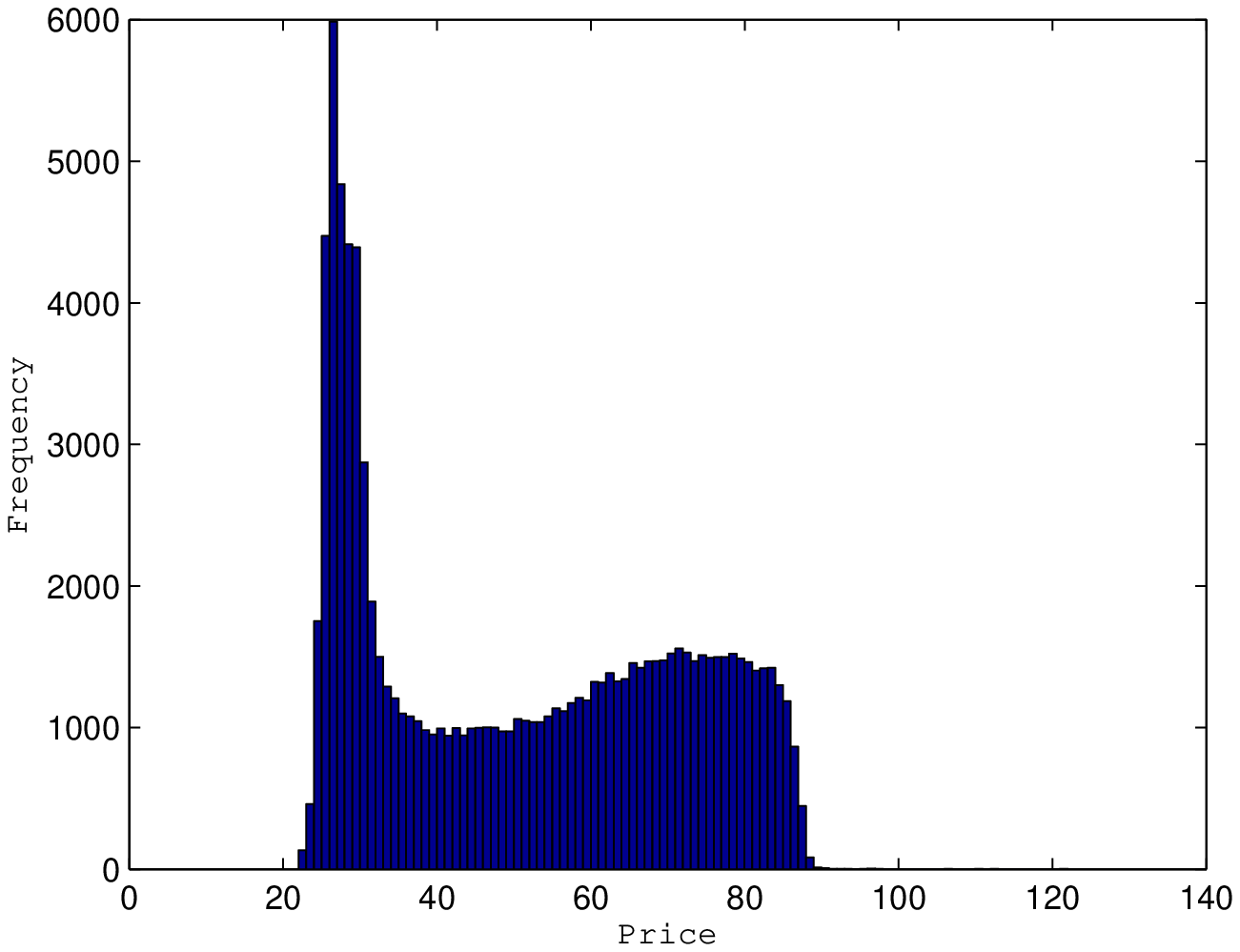}
\includegraphics[scale=0.45]{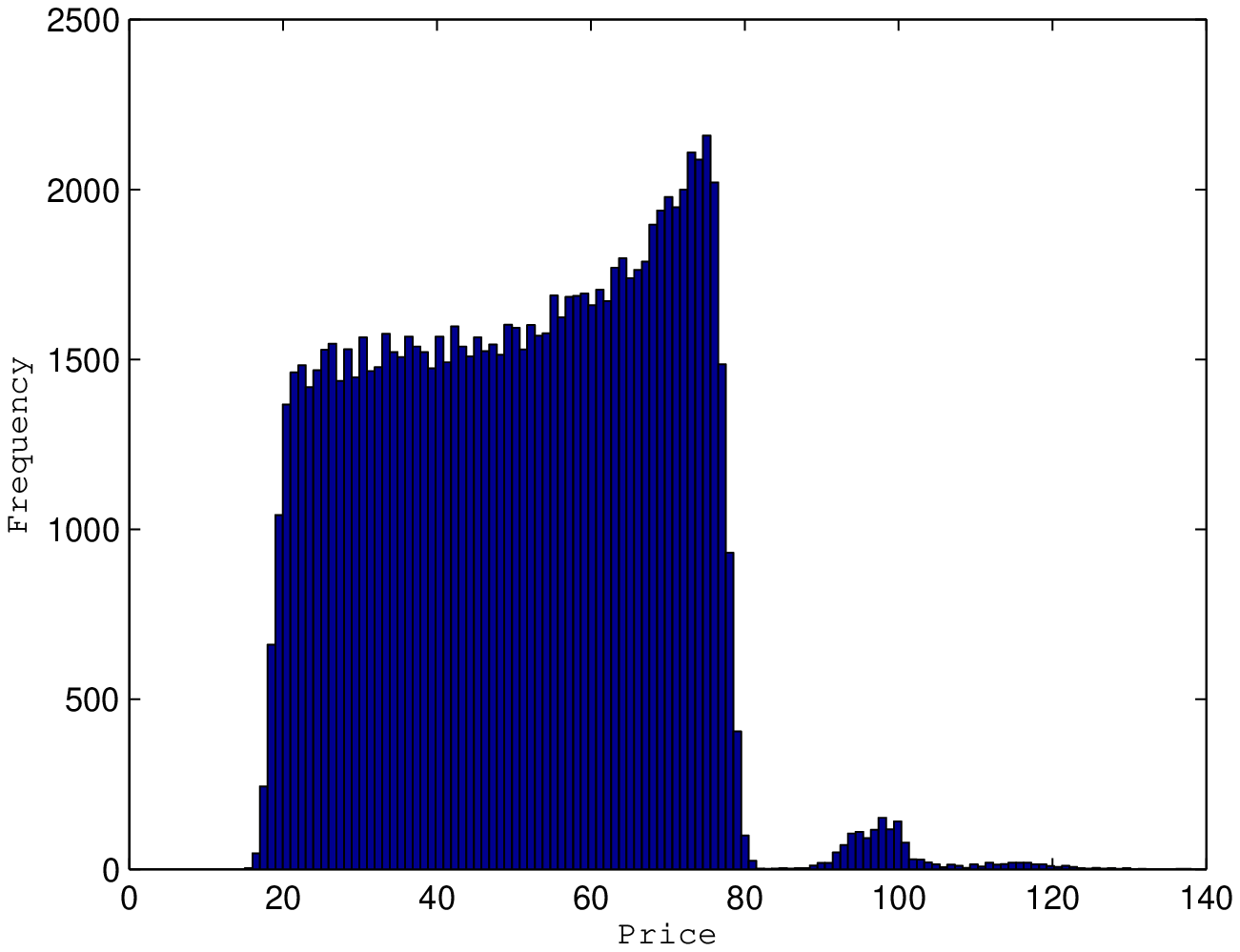}
\includegraphics[scale=0.45]{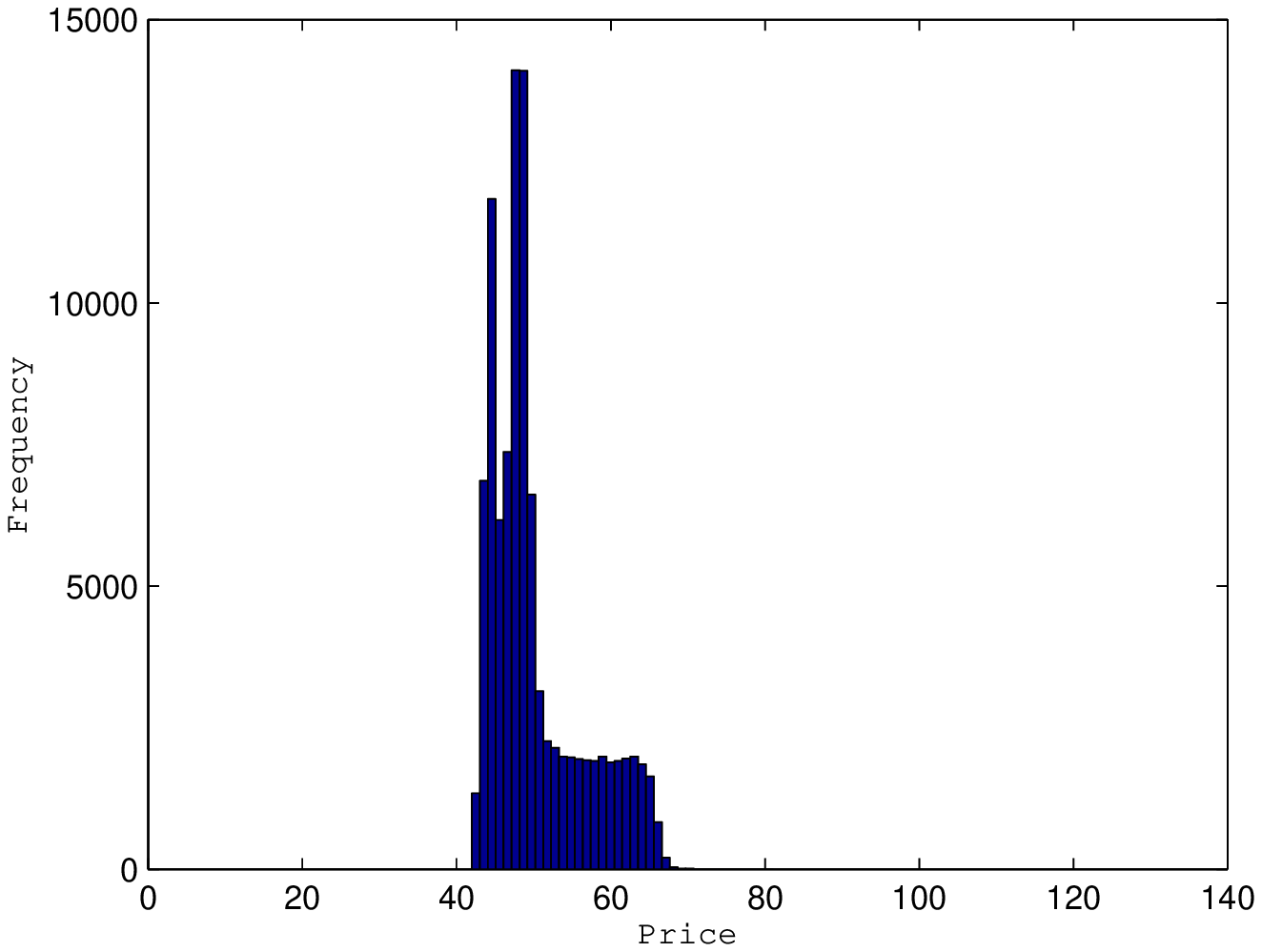}
\includegraphics[scale=0.45]{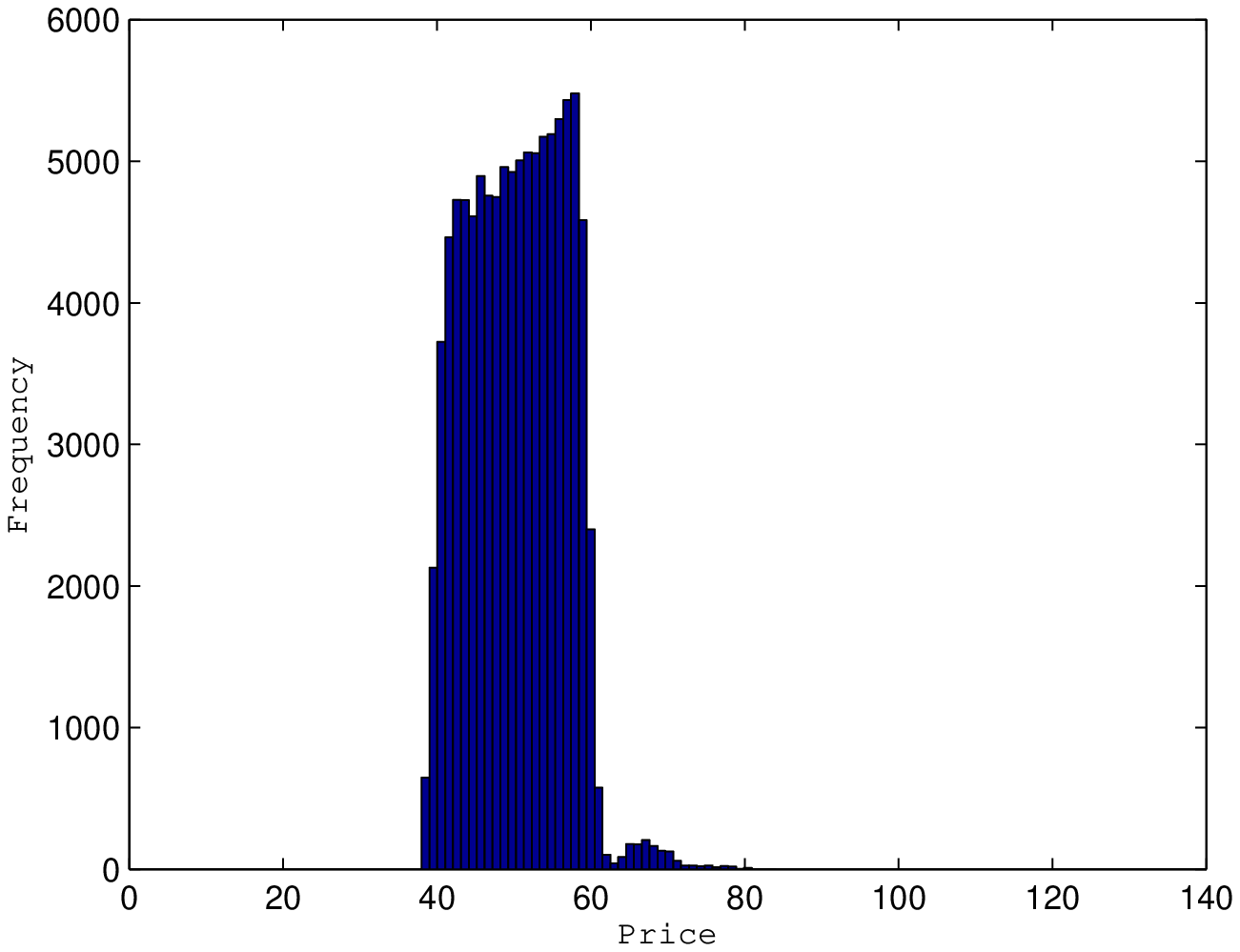}
\caption{The steady-state price distributions of the stochastic buy model with
$f_b = 0.1$, for $\beta = 30$, and $\alpha = 10$ (top left) and $\alpha = 100$
(top right), and $\beta = 10$, $\alpha = 40$ (bottom left) and $\alpha = 70$
(bottom right).}
\label{fig:Bprice}
\end{figure}

When we go to the stochastic buy model with $f_b = 0.1$, the steady-state price
distributions are so different from those of the three phases in the
deterministic model (see Figure \ref{fig:Bprice}) that we cannot
recognize the order parameters previously identified.  However, when the
stochastic buy fraction was lowered to $f_b = 0.01$, we find steady-price
distributions that closely resemble those of the deterministic model.  This
means that we can again use the height $F_0$ of the uniform sub-distribution of
below-average prices and the peak position $p_{\rm III}$ of the gaussian
sub-distribution of above-average prices as order parameters to characterize the
phase diagram of the weakly stochastic buy model.

\begin{figure}[htbp]
\centering
\includegraphics[scale=0.65]{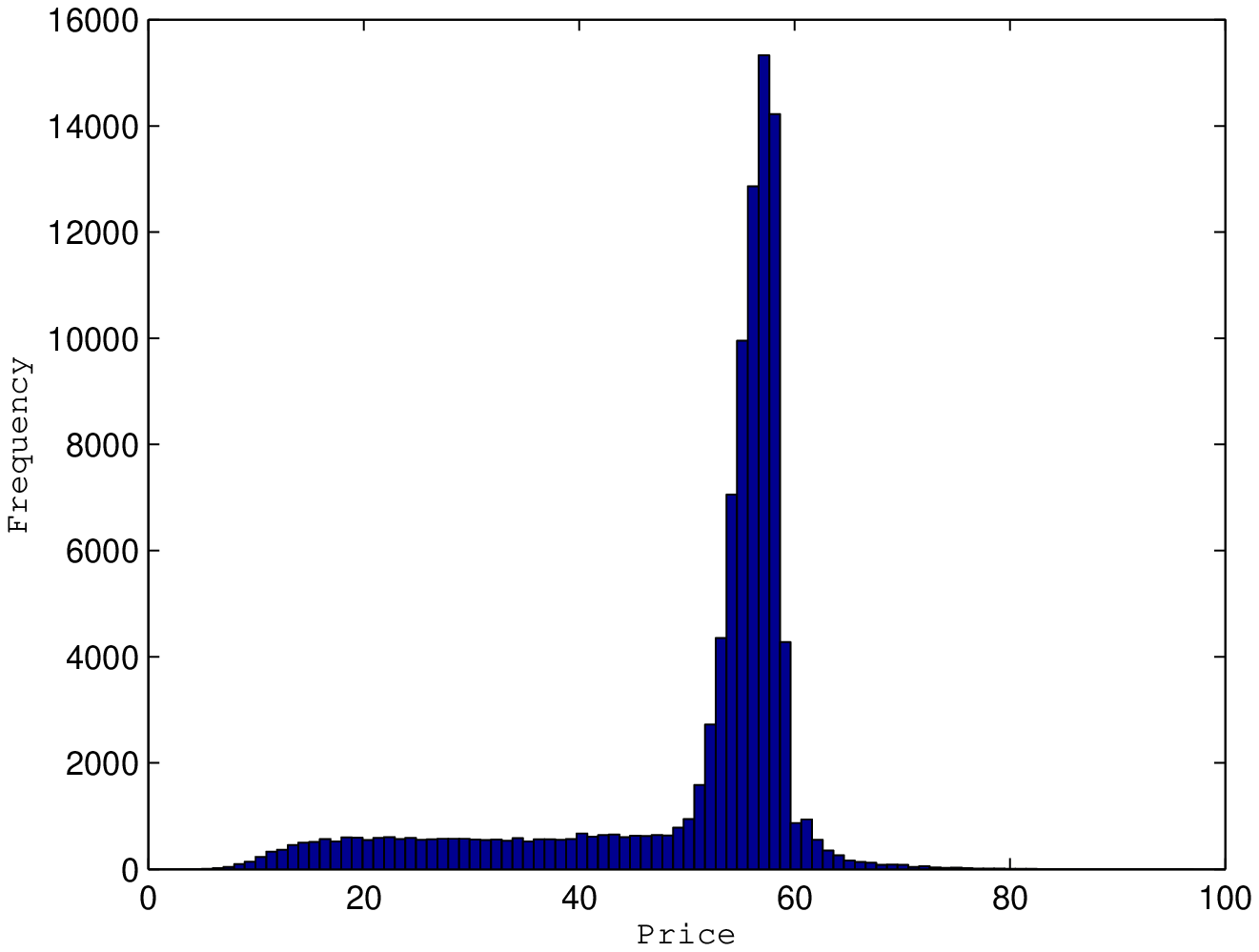}
\includegraphics[scale=0.65]{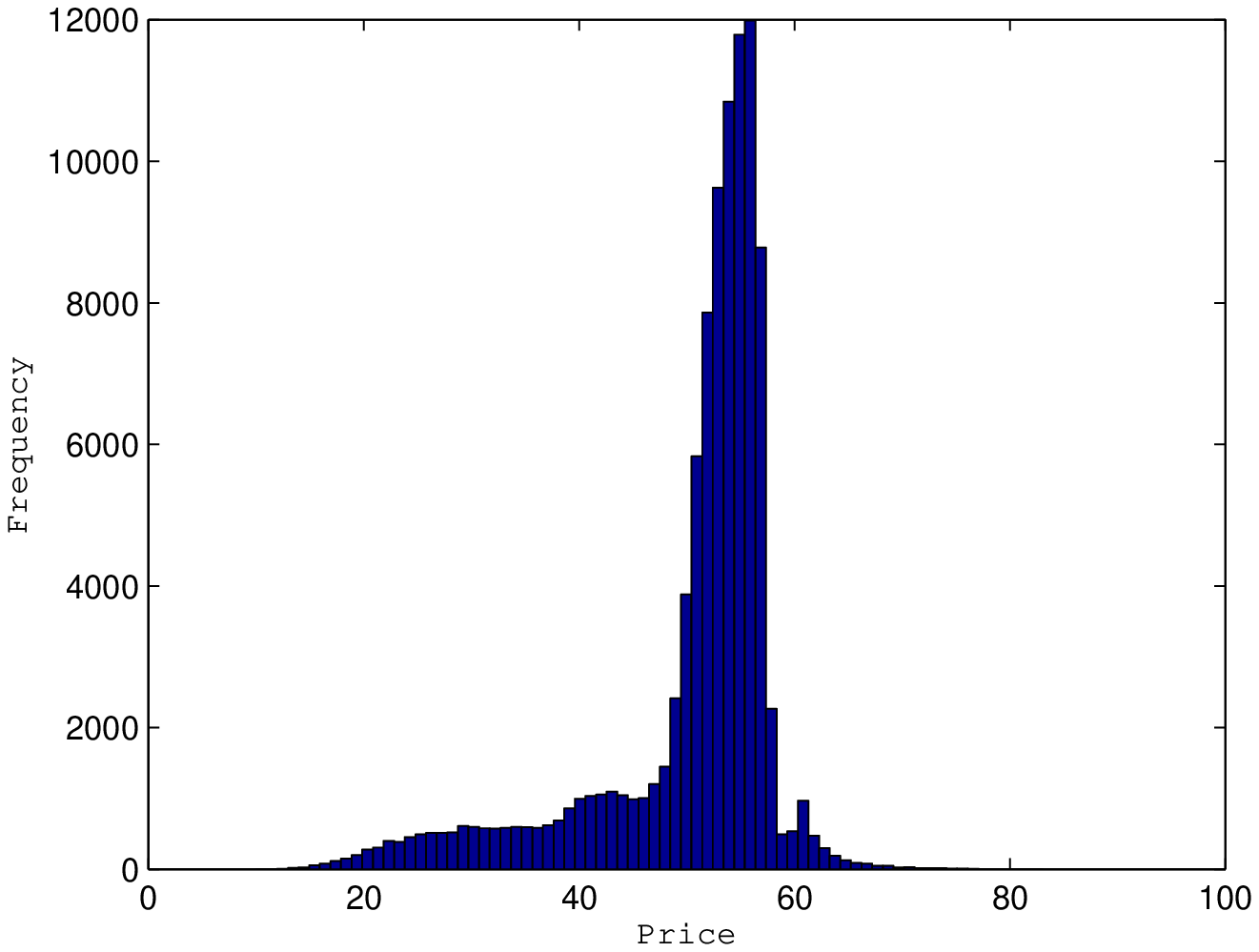}
\includegraphics[scale=0.65]{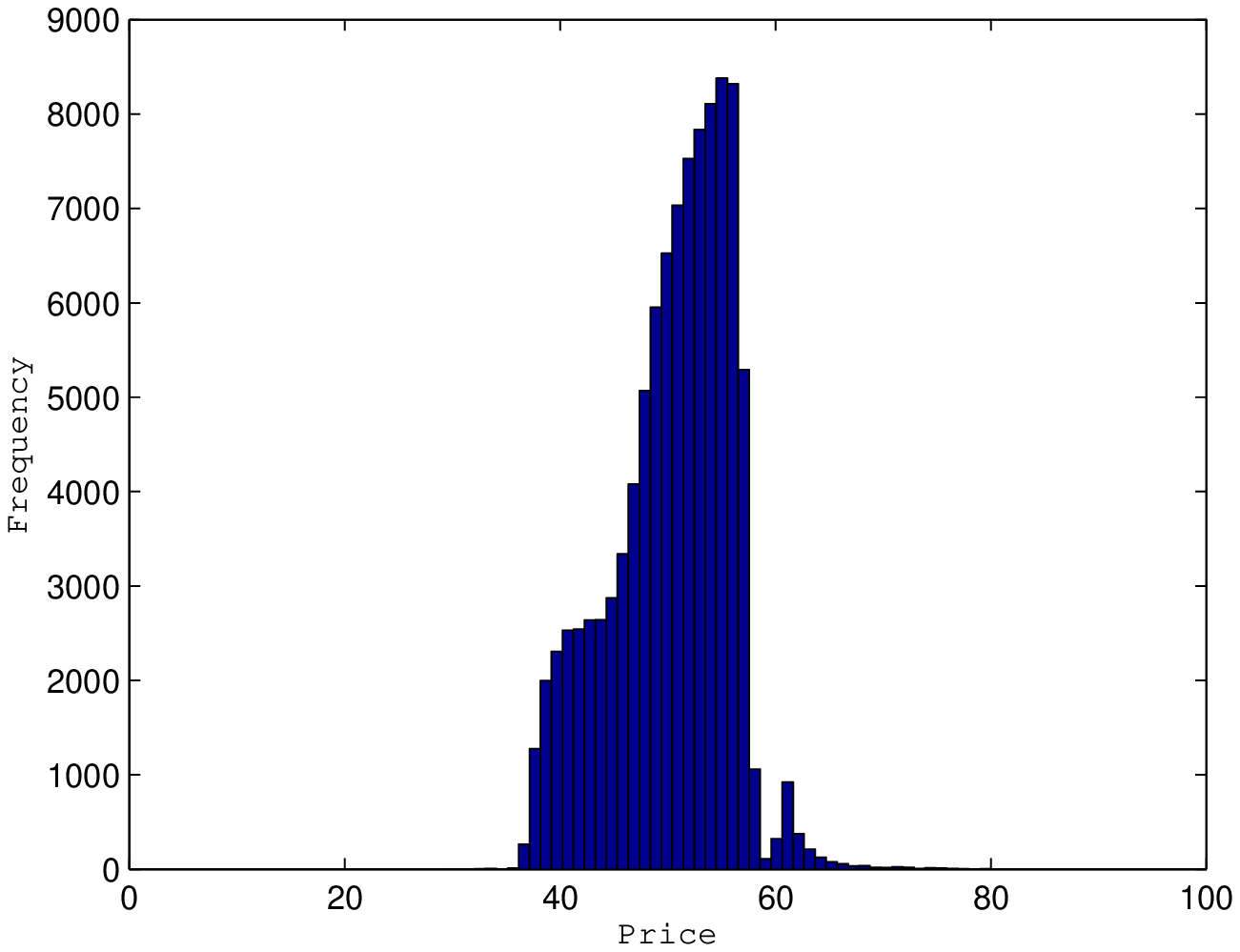}
\caption{Steady-state price distributions of the stochastic buy model with
$\alpha = 60$ and $\beta = 10$, for $f_b = 0.01$ (top), $f_b = 0.02$ (middle),
and $f_b = 0.05$ (bottom).}
\label{fig:Buniform}
\end{figure}

More importantly, we observe that the sub-distribution of below-average prices
is not uniform all the way down to $p = 0$, but develops a shoulder at $p
> 0$.  When we compare this sub-distribution for $f_b = 0.01$,
$f_b = 0.02$, and $f_b = 0.05$, as shown in Figure \ref{fig:Buniform}, we see
the shoulder price value increases as $f_b$ is increased.  This explains why we
do not find the uniform sub-distribution for $f_b = 0.1$ and higher.

\begin{figure}[htbp]
\centering
\includegraphics[scale=0.4]{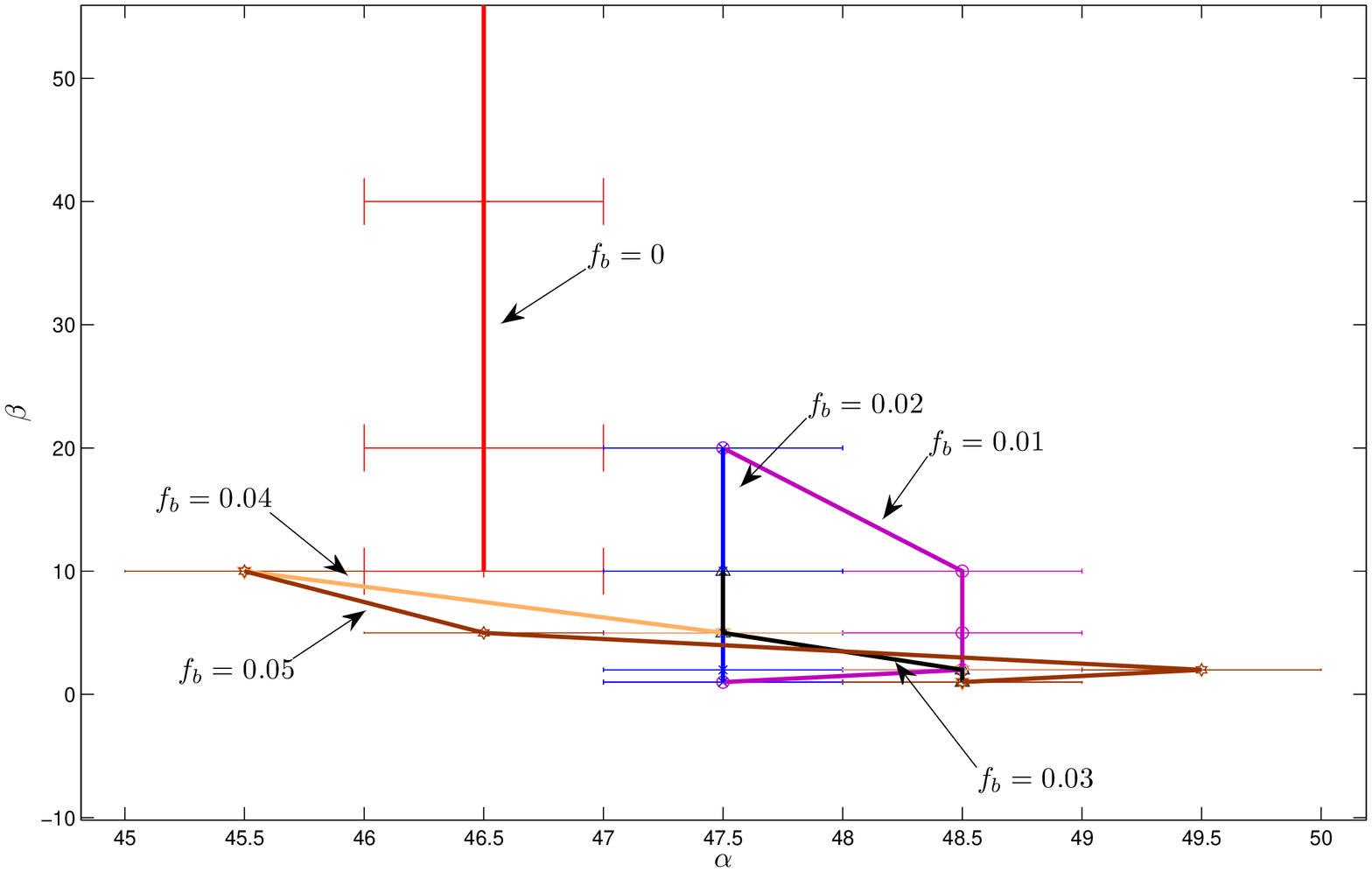}
\includegraphics[scale=0.4]{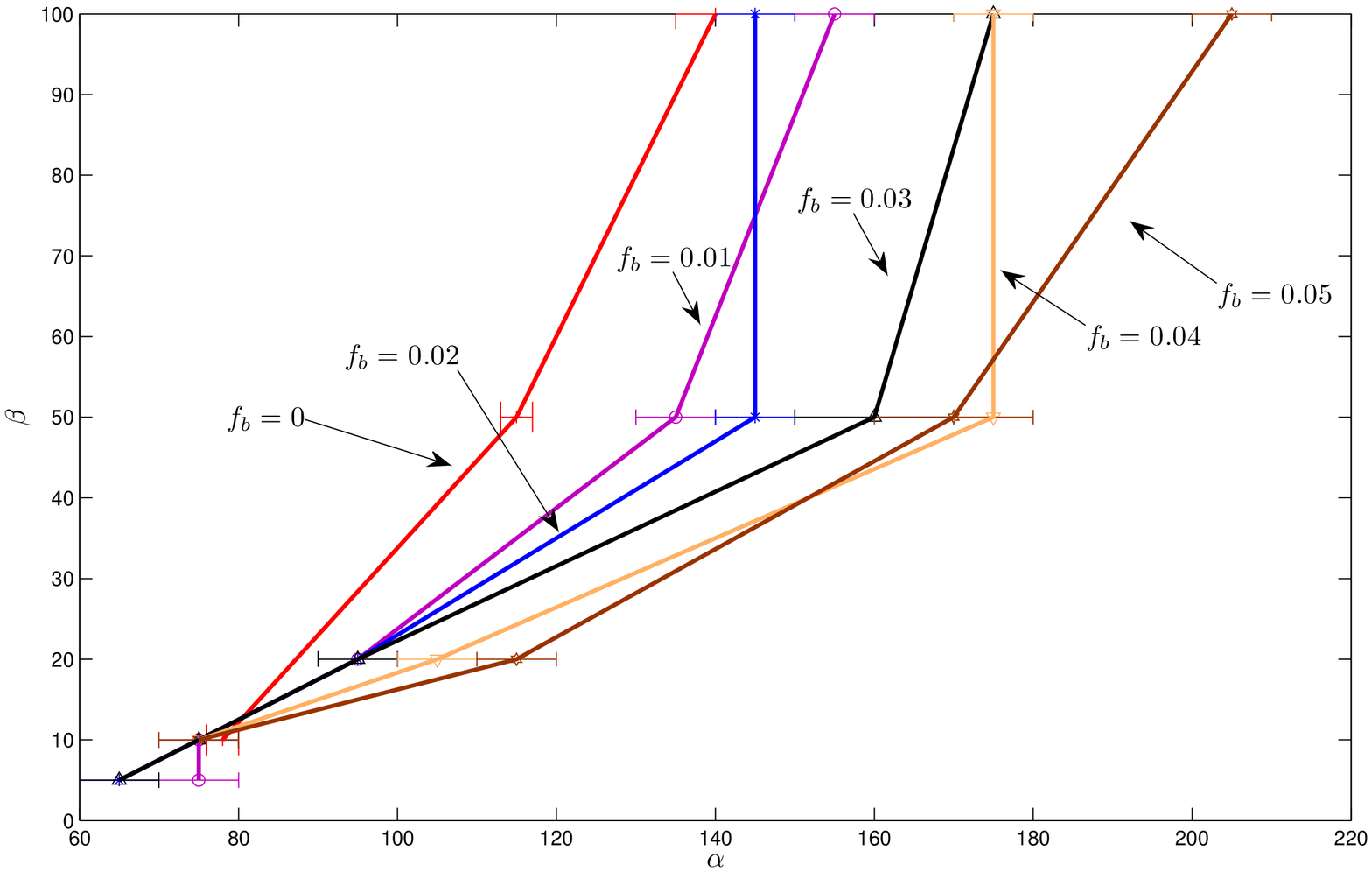}
\caption{Phase diagram of the stochastic buy model, showing how the phase
boundaries depend on the stochastic fraction $f_b$.  The error bars
$\Delta\alpha_c$ are estimated in the manner described in the caption of Figure
\ref{fig:Sphase}.}
\label{fig:Bphase}
\end{figure}

As we can see from Figure \ref{fig:Bphase}, the effects of introducing
stochasticity into the buy side of the deterministic model appears to be the
same as when it is introduced into the sell side.  The boom market phase expands
into the dead and jammed market phases.  However, random buy decisions affect
the market more strongly than random sell decisions, in the sense that a smaller
fraction of random buys produces the same effect on the phase diagram that a
larger fraction of random sells would.  There is thus strong buy-sell asymmetry
at the macroscopic level for this model, which we would not have guessed from
the microscopic rules at the agent level.

\begin{figure}[htb]
\centering
\includegraphics[scale=0.5]{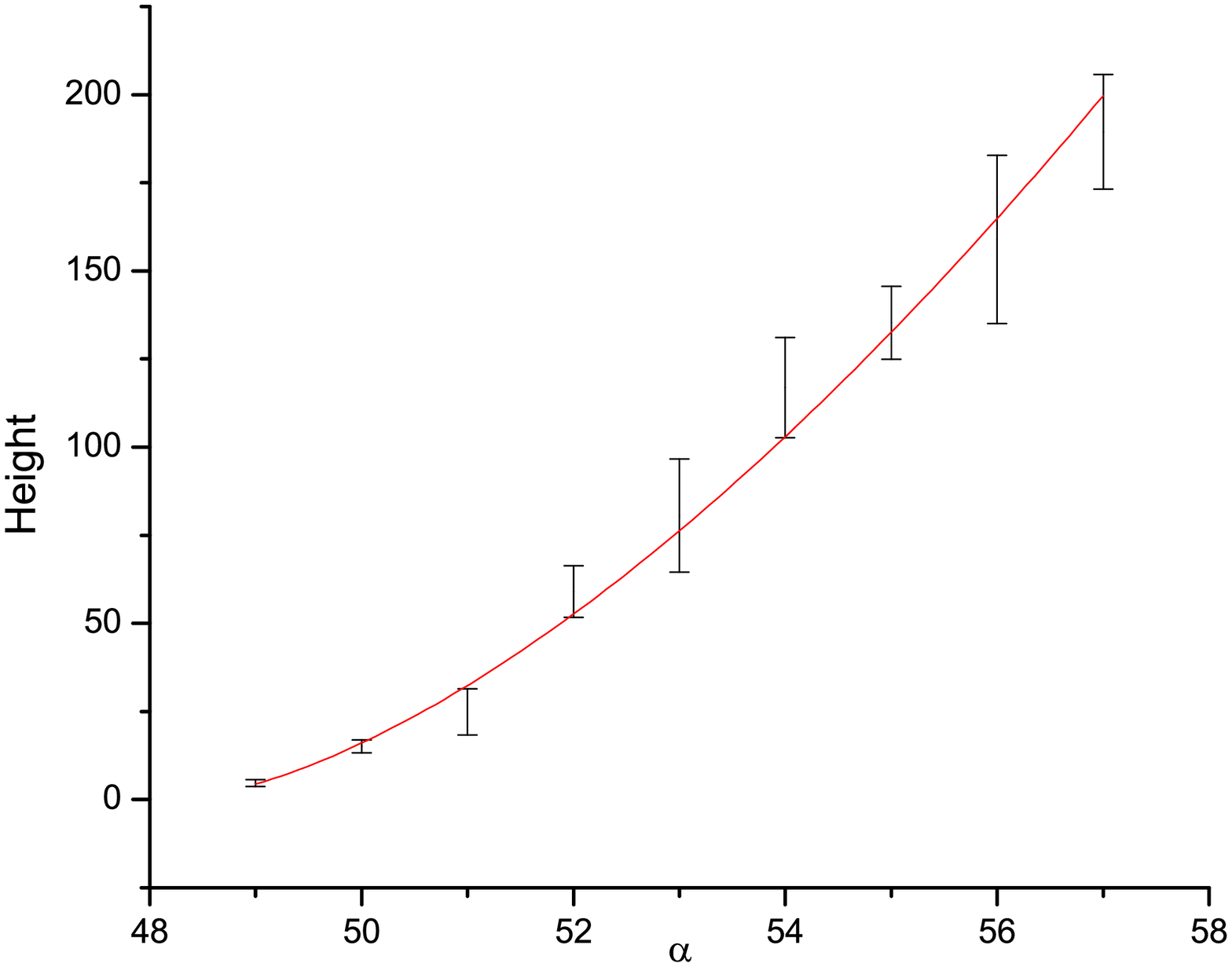}
\caption{Height $F_0$ (solid circles) of the uniform sub-distribution across the
I-II phase transition line, plotted as a function of the parameter $\alpha$, for
the stochastic buy model with $\beta = 30$ and $f_b = 0.05$ (red).  The error
bar for each data point is estimated from the uniform sub-distributions of 1000
simulations.  The solid curve is the best power-law fits of the form $F_0 = G_0
(\alpha - \alpha_c)^{\gamma}$, where we find $G_0 = 6.3 \pm 2.2$, $\alpha_c =
48.20 \pm 0.25$, $\gamma = 1.59 \pm 0.15$.}
\label{fig:Bphaseexp}
\end{figure}

Finally, we look at how the quantitative character of the I-to-II phase
transition change when we go from the deterministic model to the stochastic buy
model.  In Figure \ref{fig:Bphaseexp}, we show $F_0$ as a function of $\alpha$
for $f_b = 0.05$, fitted to a power law of the form $F_0 = G_0 (\alpha -
\alpha_c)^{\gamma}$.  Doing the same for $f_b = 0.01$ and $f_b = 0.02$, we find
the exponent going from $\gamma = 0.533 \pm 0.082$ at $f_b = 0.01$, to $\gamma =
0.90 \pm 0.13$ at $f_b = 0.02$, to $\gamma = 1.59 \pm 0.15$.  Compared to the
stochastic sell model, $\gamma$ changes very rapidly with $f_b$, with the
I-II transition becoming a crossover by $f_b \gtrsim 0.02$.

\section{Summary and Discussions}
\label{sect:summary}

In conclusion, we described in this paper a bottom-up framework for
understanding the hierarchy of complexities that emerge at different levels of
realism in agent-based modeling and simulation, and illustrated the approach
using a computational study on a simple, deterministic model and its stochastic
extensions.  In particular, we worked out the phase diagram of the deterministic
model, and identified order parameters that can be used to distinguish the three
phases: (i) dead market; (ii) boom market; and (iii) jammed market.  We find
that, as the level of stochasticity is increased, the region occupied by the
dead market phase on the phase diagram becomes smaller, and eventually
disappears when the market is sufficiently noisy.  The transition between the
dead market phase and the boom market phase was also found to be in the Ising
universality class, with critical exponent $\gamma = \frac{1}{2}$ in the
deterministic model.  We provide evidence from the trading history in the
simulations that a retarded antiferromagnetic interaction between traders,
mediated by the order book, is responsible for this critical behaviour.  When we
go from the deterministic model to the stochastic buy and sell models, this
phase transition becomes a crossover when the market becomes sufficiently noisy.

In our deterministic and stochastic sell model simulations, market rallies or
crashes were not observed.  In the context of our agent-based models, market
rallies are cooperative movements of agent trading strategies from one point
$(\alpha, \beta)$ within Phase II, to another point $(\alpha', \beta')$ within
Phase II, where $\alpha' > \alpha$.  This movement across the phase diagram can
occur under exogeneous forcing, or it can occur endogeneously through positive
feedback: if an agent learns that a fundamentalist trading strategy (governed by
the price term $\alpha/p$) is making him or her more and more money, he or she
will make his or her trading strategy even more fundamentalist by increasing
$\alpha$).  Similarly, in the context of our agent-based models, the closest
analogy to market crashes are cooperative movements of agent trading strategies
across the I-II phase transition line.  Again, this movement can have an
exogeneous origin, or it can occur endogeneously through feedback and learning.
In this study, we looked only at steady-state behaviours and worked out the
static phase diagrams of three simple models.  Ultimately, to study market rallies
and crashes, as well as temporally localized features exhibited by more
sophisticated models, we need to look more closely at the dynamics of price,
volume, and capital distributions.

In the next stage of our programme to understanding agent-based simulations, we
will simulate a more realistic model of the financial market, work out its phase
diagram, and then couple the agents to an artificial stock index (and hence more
strongly to each other), by having their trading strategies be influenced by the
stock index.  Our immediate goal for doing this would be to check for new
emergent phases, and whether it is possible to have market rallies and market
crashes \emph{without} any form of learning in the agents.  The long-term goal
is to incorporate heterogenuity, memory, and learning into our agent-based
models, and map out their diagrams of static and dynamical phases.  By
extracting the order parameters of these phases, and determining which
universality classes they belong to, we then hope to decide which market phases
are robust (appearing in a large variety of models with the same gross
structures), and which market phases are fragile (appearing only rarely within
certain models with very finely tuned parameters).  In this way, we aim to
develop a picture of the hierarchy of macroscopic phases that emerge at various
levels of realism in the agent-based models.  Ultimately, we would like to
eventually be able to say which market behaviours are robust (does not depend on
details of the models, and therefore should appear generically even in markets
with very different structures), and which behaviours are fragile (depends
sensitively on certain model details, and therefore appear rarely in real-world
markets).

\section*{Acknowledgments}

This work is supported by startup grant SUG 19/07 from the Nanyang
Technological University.


\begin{thebibliography}{99}

\bibitem{Palmer1994PhysicaD75p264} R. G. Palmer, W. B. Arthur, J. H.
Holland, B. LeBaron, and P. Tayler, ``Artificial Economic Life: A
Simple Model of a Stockmarket'', Physica D, vol. 75, pp. 264--274,
1994.

\bibitem{Epstein1999Complexity4p41} J. M. Epstein, ``Agent-Based
Computational Models and Generative Social Science'', Complexity, vol.
4, no. 5, pp. 41--60, 1999.

\bibitem{Marsili2007EurPhysJB55p169} M. Marsili, ``Toy Models and
Stylized Realities'', The European Physical Journal B, vol. 55, pp.
169--173, 2007.

\bibitem{Farmer2002IndCorpChange11p895} J. D. Farmer, ``Market Force,
Ecology and Evolution'', Industrial and Corporate Change, vol. 11, no.
5, pp. 895--953, 2002.

\bibitem{LeBaron2000JEconDynCont24p679} B. LeBaron, ``Agent-Based
Computational Finance: Suggested Readings and Early Research'',
Journal of Economic Dynamics \& Control, vol. 24, pp. 679--702, 2000.

\bibitem{LeBaron2001QuantFin1p254} B. LeBaron, ``A Builder's Guide to
Agent-Based Financial Markets'', Quantitative Finance, vol. 1, pp.
254--261, 2001.

\bibitem{Tesfatsion2002ArtificialLife8p55} L. Tesfatsion,
``Agent-Based Computational Economics: Growing Economies From Botton
Up'', Artificial Life, vol. 8, no. 1, pp. 55--82, 2002.

\bibitem{HCE2006} \textsl{Handbook of Computational Economics, Volume
2: Agent-Based Computational Economics}, edited by L. Tesfatsion and
K. Judd, North-Holland, 2006.

\bibitem{Levy2000MicroSimulFinMkt} M. Levy, H. Levy, and S. Solomon,
\textsl{Microscopic Simulation of Financial Markets}, Academic Press,
2000.

\bibitem{Mantegna2000IntroEconophysics} R. N. Mantegna and H. E.
Stanley, \textsl{An Introduction to Econophysics: Correlations and
Complexity in Finance}, Cambridge University Press, 2008.

\bibitem{Lux1999Nature297p498} T. Lux and M. Marchesi, ``Scaling and
Criticality in a Stochastic Multi-Agent Model of a Financial Market'',
Nature, vol. 397, no. 6719, pp. 498--500, 1999.

\bibitem{Lux2000IntJTheorApplFin3p675} T. Lux and M. Marchesi,
``Volatility Clustering in Financial Markets: A Microsimulation of
Interacting Agents'', International Journal of Theoretical and Applied
Finance, vol. 3, no. 4, pp. 675--702, 2000.

\bibitem{Raberto2001PhysicaA299p319} M. Raberto, S. Cincotti, S. M.
Focardi, and M. Marchesi, ``Agent-Based Simulation of a Financial
Market'', Physica A, vol. 299, pp. 319--327, 2001.

\bibitem{LeBaron1999JEconDynCont23p1487} B. LeBaron, W. B. Arthur, and
R. Palmer, ``Time Series Properties of An Artificial Stock Market'',
Journal of Economic Dynamics \& Control, vol. 23, pp. 1487--1516,
1999. 

\bibitem{Bak1997PhysicaA246p430} P. Bak, M. Paczuski, and M. Shubik,
``Price Variations in a Stock Market with Many Agents'', Physica A,
vol. 246, pp. 430--453, 1997.

\bibitem{Caldarelli1997EuroPhysLett40p479} G. Caldarelli, M. Marsili,
and Y.-C. Zhang, ``A Prototype Model of Stock Exchange'', Europhysics
Letters, vol. 40, no. 5, pp. 479--484, 1997.

\bibitem{Sato1998PhysicaA250p231} A.-H. Sato and H. Takayasu,
``Dynamic Numerical Models of Stock Market Price: From Microscopic
Determinism to Macroscopic Randomness'', Physica A, vol. 250, pp.
231--252, 1998.

\bibitem{Chowdhury1999EurPhysJB8p477} D. Chowdhury and D. Stauffer,
``A Generalized Spin Model of Financial Markets'', The European
Physical Journal B, vol. 8, pp. 477--482, 1999.

\bibitem{Cont2000MacroEconDyn4p170} R. Cont and J.-P. Bouchaud, ``Herd
Behavior and Aggregate Fluctuations in Financial Markets'',
Macroeconomic Dynamics, vol. 4, pp. 170--196, 2000.

\bibitem{Maslov2000PhysicaA278p571} S. Maslov, ``Simple Model of a
Limit Order-Driven Market'', Physica A, vol. 278, pp. 571--578, 2000.

\bibitem{Iori2002JEconBehavOrg49p269} G. Iori, ``A Microsimulation of
Traders Activity in the Stock Market: The Role of Heterogenuity,
Agents' Interactions and Trade Frictions'', Journal of Economic
Behavior \& Organization, vol. 49, pp. 269--285, 2002.

\bibitem{Giardina2003EurPhysJB31p421} I. Giardina and J.-P. Bouchaud,
``Bubbles, Crashes and Intermittency in Agent Based Market Models'',
The European Physical Journal B, vol. 31, pp. 421--437, 2003.

\bibitem{Giardina2003PhysicaA324p6} I. Giardina and J.-P. Bouchaud,
``Volatility Clustering in Agent Based Market Models'', Physica A,
vol. 324, pp. 6--16, 2003.  

\bibitem{Moukarzel2007EurPhysJSpecTop143p75} C. F. Moukarzel, S.
Gon\c{c}alves, J. R. Iglesias, M. Rodr\'{\i}guez-Achach, and R.
Huerta-Quintanilla, ``Wealth Condensation in a Multiplicative Random
Asset Exchange Model'', The European Physical Journal: Special Topics,
vol. 143, pp. 75--79, 2007.

\bibitem{Gauvin2009socph09034694} L. Gauvin, J. Vannimenus, and J.-P.
Nadal, ``Phase Diagram of a Schelling Segregation Model'', The European Physical
Journal B, vol. 70, pp. 293--304, 2009.

\bibitem{Binney1992} J. J. Binney, N. J. Dowrick, A. J. Fisher, and M. E. J.
Newman, \textsl{The Theory of Critical Phenomena: An Introduction to the
Renormalization Group}, Oxford University Press, 1992.

\bibitem{Johansen1999EurPhysJB9p167} A. Johansen and D. Sornette,
``Modeling the Stock Market Prior to Large Crashes'', The European
Physical Journal B, vol. 9, no. 1, pp. 167--174, 1999.

\bibitem{Johansen2000IntJTheorApplFin3p219} A. Johansen, O. Ledoit,
and D. Sornette, ``Crashes As Critical Points'', International Journal
of Theoretical and Applied Finance, vol. 3, no. 2, pp. 219--255, 2000.

\end{thebibliography}
\end{document}